\documentclass[superscriptaddress, prd, aps,amsmath,amssymb,showpacs,showkeys, onecolumn]{revtex4-2}
\usepackage[dvips]{graphicx,color}
\usepackage{subfig}
\usepackage{times}
\usepackage{xcolor}
\usepackage[%
  colorlinks=true,
  urlcolor=blue,
  linkcolor=red,
  citecolor=blue
]{hyperref}
\usepackage{orcidlink}

\begin{document}
\title{Extended Thermodynamics and Throttling Process of Charged AdS Black Holes in ModMax-dRGT Massive Gravity with Sharma-Mittal Entropy}

\author{Naba Jyoti Gogoi\orcidlink{0000-0002-1693-9501}}
\email[Email: ]{gogoin799@gmail.com}

\affiliation{Department of Physics, Sibsagar University, Joysagar 785665, Assam, India}

\affiliation{Research Center of Astrophysics and Cosmology, Khazar University, Baku, AZ1096, 41 Mehseti Street, Azerbaijan}

 \author{Hassan Hassanabadi}
\email[Email: ]{hha1349@gmail.com}
\affiliation{Department of Physics, Faculty of Science, University of Hradec Kr$\acute{a}$lov$\acute{e}$, Rokitansk$\acute{e}$ho 62, 500 03 Hradec Kr$\acute{a}$lov$\acute{e}$, Czechia.}
\affiliation{Department of Physics and Electronics, Khazar University, 41 Mahsati Str, 1096 Baku, Azerbaijan}
\affiliation{Al-Farabi Kazakh National University, Al-Farabi ave. 71, 050040 Almaty, Kazakhstan}

\author{Himasri Pinapothu \orcidlink{0009-0000-2312-4610}}%
 \email[Email: ]{pinapothuhimasri@gmail.com}
\affiliation{%
 Department of Physics, Andhra Loyola College, Vijayawada, Andhra Pradesh, India 520008}

\author{Dhruba Jyoti Gogoi \orcidlink{0000-0002-4776-8506}}%
 \email[Email: ]{moloydhruba@yahoo.in}
\affiliation{Department of Physics, Madhabdev University, Narayanpur, Lakhimpur 784164, Assam, India}
\affiliation{{Jadara Research Center, Jadara University, Irbid 21110, Jordan}}

\begin{abstract}
We investigate the extended thermodynamics, including the Joule-Thomson expansion and $P-V$ criticality, of a four-dimensional charged anti-de Sitter (AdS) black hole within the combined framework of ModMax nonlinear electrodynamics and dRGT-like massive gravity. Operating in the extended phase space and employing the generalised Sharma-Mittal entropy to account for non-extensive statistical correlations, we derive exact analytical expressions for the modified Hawking temperature, specific heat, Joule-Thomson coefficient, and the equation of state. Our analysis of the throttling process reveals that the conformal nonlinearities of the ModMax field ($\gamma$) expand the physically accessible cooling domain by shifting the inversion transition to smaller horizon radii. While the Sharma-Mittal parameters ($\delta$, $R$) critically govern local thermodynamic stability and the inversion radius, the global inversion phase boundary remains fundamentally dictated by the massive graviton background. Furthermore, an analysis of the Gibbs free energy uncovers a van der Waals-like first-order phase transition characterized by a distinct swallow-tail structure. We observe a clear physical decoupling in the critical regime: ModMax nonlinearities modify the critical phase boundary by suppressing electromagnetic interactions, Sharma-Mittal parameters dictate the relative thermal stability of competing phases, and massive gravity governs the overarching macroscopic phase landscape. These results highlight the sensitivity of thermodynamic phase phenomena as robust diagnostic tools for distinguishing nonlinear and non-extensive modifications to black hole physics.
\end{abstract}

\keywords{Black Hole Thermodynamics; ModMax-dRGT-like massive gravity; Entropy; Joule Thomson expansion}

\maketitle

\section{Introduction}
\label{sec1}

The formulation of black hole thermodynamics has established a profound and enduring connection between general relativity, quantum mechanics, and statistical physics. Initiated by the seminal discoveries of black hole entropy and Hawking radiation \cite{Bekenstein:1973ur, Hawking:1975vcx}, black holes are no longer viewed merely as astronomical endpoints, but as fundamental theoretical laboratories. A critical advancement in this domain is the extended phase space formalism, wherein the negative cosmological constant $\Lambda$ is dynamically interpreted as a thermodynamic pressure, $P=-\frac{\Lambda}{8\pi}$, and the total mass $M$ of the black hole is identified with the thermodynamic enthalpy \cite{Kastor:2009wy}. This paradigm shift has revealed that anti-de Sitter (AdS) black holes exhibit a rich phase structure strikingly analogous to conventional thermodynamic fluids, encompassing van der Waals-like phase transitions, triple points, and heat engine cycles \cite{Hawking:1982dh,Mann:2025xrb,Altamirano:2013ane,Li:2025lrq,Kubiznak:2012wp,Gogoi:2021syo,Gogoi:2023xzy,Gogoi:2023qni,Gogoi:2023wih,Gogoi:2024akv}.

A fascinating application of this extended formalism is the Joule-Thomson (JT) expansion of black holes. In classical thermodynamics, the JT process characterizes the temperature change of a real gas as it undergoes an isenthalpic (constant enthalpy) expansion through a porous plug. For black holes, this translates to a constant-mass ($dM=0$) expansion process driven by a decrease in the AdS pressure. The central diagnostic of this process is the Joule-Thomson coefficient, which dictates whether the black hole undergoes cooling or heating. The boundary separating these two thermodynamic regimes is defined by the inversion curve. Since its initial exploration in charged AdS black holes \cite{Okcu:2016tgt}, the JT expansion has become a standard tool to probe the thermodynamic microstructure of black holes across various modified gravity and matter theories \cite{Gogoi:2023ntt,Chabab:2018zix,Liu:2024zzr,Yasir:2023aup,Chaudhary:2022sfg,Alessa:2026oxg,Ahmed:2026zax,Sekhmani:2022hir,Rajani:2020mdw, Ahmed:2026zax, Qi:2026oxm, Fatima:2025aqb, Media:2025gvl, Liu:2025ulr, Liu:2024zzr, Shahzad:2024cvs, Shahzad:2024ycq, Alipour:2024xxs, Zhang:2024fxj, Li:2023mql, Du:2023kgj, Guo:2023plg, Sekhmani:2023est, Zhang:2022cdp, Assrary:2022uiu, Cao:2022hmd, Xing:2021gpn, Meng:2021cgb, Zhang:2021kha, Ghaffarnejad:2020nhb, Jawad:2020hju, Feng:2020swq, Meng:2020csd, Guo:2020qxy, Hegde:2020cdm, Nam:2020gud, Rajani:2020mdw, Lan:2019kak, Cisterna:2018jqg, RizwanCL:2018cyb, Mo:2018qkt, Mo:2018rgq, Ghaffarnejad:2018exz}.

Concurrently, theoretical cosmology and high-energy physics have been driven to explore extensions of general relativity and standard electrodynamics to resolve observational tensions and singularities. On the gravity side, the ghost-free dRGT massive gravity \cite{deRham:2010kj,deRham:2011rn} endows the graviton with a non-zero mass $m_g$, offering a natural mechanism for the late-time acceleration of the universe. The dRGT-like extension, utilising a singular reference metric \cite{Zhang:2015nwy}, yields exact black hole solutions characterised by massive parameters that significantly alter the spacetime structure and thermodynamic phase space \cite{Cai:2015fia}.

On the matter side, nonlinear electrodynamics (NLE) was originally proposed to cure the infinite self-energy of point charges. While the Born-Infeld model is the most renowned NLE, a novel formulation known as Modified Maxwell (ModMax) electrodynamics has recently garnered significant attention \cite{PhysRevD.102.121703}. ModMax is uniquely distinguished as the only continuous one-parameter nonlinear extension of classical electrodynamics that exactly preserves both conformal invariance and electromagnetic duality. The presence of the dimensionless ModMax parameter $\gamma$ introduces conformal non-linearities that fundamentally modify the electromagnetic field without breaking the underlying symmetries of Maxwell's theory.

Motivated by these developments, this paper investigates the Joule-Thomson expansion of charged AdS black holes within the combined framework of ModMax nonlinear electrodynamics and dRGT-like massive gravity. While the thermodynamic properties and phase transitions of these black holes have been recently explored, the throttling process in a spacetime governed by massive gravitons and ModMax fields remains largely unaddressed. Our objective is to rigorously derive the Joule-Thomson coefficient, the inversion temperature, and the isenthalpic curves, thereby unearthing how the ModMax parameter and massive gravity couplings shift the inversion boundaries and alter the heating-cooling domains of the black hole.

The remainder of this paper is organised as follows. In Sec. \ref{sec2}, we briefly review the exact black hole solution and its thermodynamic quantities in the extended phase space of ModMax-dRGT-like massive gravity. In Sec. \ref{sec3}, we formulate the Joule-Thomson expansion, derive the inversion pressure and temperature, and analyse the resulting isenthalpic curves. $P - V$ criticality and phase structure have been investigated in Sec.\ref{sec:pv-criticality}. Sec. \ref{sec:gibbs-free-energy} deals with Gibbs free energy and global stability of the system. Finally, our concluding remarks and physical interpretations are presented in Sec. \ref{sec4}. Throughout this work, we adopt geometrodynamic units with $G=\hbar=c=1$.

\section{Black hole solution and extended thermodynamics in ModMax-dRGT-like massive gravity}
\label{sec2}

To investigate the throttling process, we first discuss the corresponding thermodynamic quantities of the black hole in dRGT-like massive gravity coupled with ModMax nonlinear electrodynamics in the presence of a negative cosmological constant $\Lambda$ introduced in Ref. \cite{EslamPanah:2025bfh}. The static, spherically symmetric line element takes the standard form:
\begin{equation}
    ds^{2}=-\psi(r)dt^{2}+\frac{dr^{2}}{\psi(r)}+r^{2}(d\theta^{2}+\sin^{2}\theta d\phi^{2}),
    \label{metric}
\end{equation}
where the metric function $\psi(r)$ encapsulates the influences of both the massive graviton and the nonlinear ModMax field. By solving the corresponding field equations, the metric function is obtained as \cite{EslamPanah:2025bfh}:
\begin{equation}
    \psi(r)=1-\frac{m_{0}}{r}-\frac{\Lambda}{3}r^{2}+\frac{q^{2}e^{-\gamma}}{r^{2}}+m_{g}^{2}C\left( \frac{c_{1}r}{2}+c_{2}C\right),
    \label{sol}
\end{equation}
where $m_{0}$ is a constant related to the total mass of the black hole, $q$ is the integration constant associated with the electric charge, and $\gamma \geq 0$ is the dimensionless ModMax parameter. The terms proportional to $m_{g}^{2}$ originate from the massive graviton, where $c_{1}$ and $c_{2}$ are arbitrary constants of the massive potential, and $C$ is a positive constant originating from the spatial reference metric. Notably, in the limit $\gamma \to 0$, the ModMax field reduces to standard Maxwell electrodynamics, and for $m_g \to 0$, one recovers the standard Reissner-Nordstr\"om-AdS (RN-AdS) solution.

In the extended phase space formalism of black hole thermodynamics, the cosmological constant is reinterpreted as a dynamical thermodynamic pressure $P$:
\begin{equation}
    P=-\frac{\Lambda}{8\pi}.
    \label{pressure_def}
\end{equation}
Correspondingly, the total mass $M = m_{0}/2$ of the black hole is no longer interpreted strictly as internal energy, but rather as the thermodynamic enthalpy of the system, $M \equiv H$. The event horizon of the black hole, $r_+$, is determined by the largest positive root of $\psi(r_+) = 0$. By substituting Eq.~\eqref{pressure_def} into Eq.~\eqref{sol} and solving $\psi(r_+) = 0$ for $M$, the enthalpy is expressed as:
\begin{equation}
    M = \frac{r_{+}}{2}+\frac{4\pi P r_{+}^{3}}{3}+\frac{q^{2}e^{-\gamma}}{2r_{+}}+\frac{m_{g}^{2}Cr_{+}}{2}\left( \frac{c_{1}r_{+}}{2}+c_{2}C\right).
    \label{massEq}
\end{equation}

The thermodynamic volume $V$ conjugate to the pressure $P$ is obtained via the standard thermodynamic relation $V = \left(\frac{\partial M}{\partial P}\right)_{S, Q, c_i}$, which yields:
\begin{equation}
    V = \frac{4\pi r_{+}^{3}}{3}.
    \label{volume}
\end{equation}
This confirms that the thermodynamic volume geometrically coincides with the spatial volume of a sphere of radius $r_+$.

The Hawking temperature of the black hole, defined via the surface gravity $\kappa$ evaluated at the event horizon ($T_H = \frac{\kappa}{2\pi} = \frac{\psi'(r_+)}{4\pi}$), is calculated as:
\begin{equation}
    T_H = 2P r_{+} + \frac{1}{4\pi}\left( \frac{1}{r_{+}} - \frac{q^{2}e^{-\gamma}}{r_{+}^{3}} \right) + \frac{m_{g}^{2}C}{4\pi}\left( c_{1} + \frac{c_{2}C}{r_{+}} \right).
    \label{tempval}
\end{equation}
The presence of the ModMax parameter $\gamma$ explicitly exponentially suppresses the electromagnetic contribution to the temperature, whereas the massive gravity parameters $(c_1, c_2)$ introduce scale-dependent modifications to the horizon's thermal output. 

By rearranging Eq.~\eqref{tempval}, we can obtain the geometric equation of state $P(V, T_H)$ for the ModMax-dRGT-like black hole:
\begin{equation}
    P = \frac{T_H}{2r_{+}} - \frac{1}{8\pi r_{+}^{2}} + \frac{q^{2}e^{-\gamma}}{8\pi r_{+}^{4}} - \frac{m_{g}^{2}C}{8\pi r_{+}}\left( c_{1} + \frac{c_{2}C}{r_{+}} \right).
    \label{EoS}
\end{equation}
This equation of state acts as the gravitational analogue to the van der Waals fluid equation. It provides the foundation necessary for investigating isenthalpic processes, specifically the Joule-Thomson expansion, which we will address in the subsequent section.

\section{Sharma-Mittal Entropy and Joule-Thomson Expansion}
\label{sec3}

\subsection{Generalised Entropy and Modified Temperature}
In systems involving exotic matter fields, such as the conformal non-linearities of ModMax electrodynamics or the long-range interactions induced by massive gravitons, the standard Boltzmann-Gibbs entropy may not provide a complete thermodynamic description. To capture potential non-extensive effects, we employ the Sharma-Mittal (SM) generalised entropy, defined as \cite{SayahianJahromi:2018irq}:
\begin{equation}
    \mathcal{S} = \frac{1}{R} \left[ \left(1 + \delta S_T \right)^{\frac{R}{\delta}} - 1 \right],
    \label{SM_def}
\end{equation}
where $S_T = \pi r_+^2$ is the standard Tsallis/Bekenstein-Hawking entropy, and $R$ and $\delta$ are free non-extensive parameters. This two-parameter framework is highly versatile, reducing to R\'enyi entropy as $\delta \to 0$, Tsallis entropy as $R \to \delta$, and recovering the standard area law for $R \to 0$ and $\delta \to 0$.

The modified Hawking temperature, $T_{SM}$, is obtained via the first law of thermodynamics, $T = (\partial M / \partial \mathcal{S})_P$. Applying the chain rule $T_{SM} = (\partial M / \partial r_+) / (\partial \mathcal{S} / \partial r_+)$, we find:
\begin{equation}
    T_{SM} = T_0 (r_+, P) \cdot \left(1 + \delta \pi r_+^2 \right)^{1 - \frac{R}{\delta}},
    \label{TSM}
\end{equation}
where $T_0(r_+, P)$ is the standard Hawking temperature given in Eq. (\ref{tempval}). By defining the SM modification factor as $f(r_+) \equiv (1 + \delta \pi r_+^2)^{1 - R/\delta}$, the equation of state becomes:
\begin{equation}
    P = \frac{T_{SM} f(r_+)^{-1}}{2r_{+}} - \frac{1}{8\pi r_{+}^{2}} + \frac{q^{2}e^{-\gamma}}{8\pi r_{+}^{4}} - \frac{m_{g}^{2}C}{8\pi r_{+}}\left( c_{1} + \frac{c_{2}C}{r_{+}} \right).
    \label{EoS_SM}
\end{equation}

\subsection{Joule-Thomson Expansion}
The Joule-Thomson expansion describes an isenthalpic process ($dM=0$). The thermodynamic response is governed by the Joule-Thomson coefficient $\mu$:
\begin{equation}
    \mu = \left( \frac{\partial T_{SM}}{\partial P} \right)_M = \frac{1}{C_P} \left[ T_{SM} \left( \frac{\partial V}{\partial T_{SM}} \right)_P - V \right],
\end{equation}
where $C_P = T_{SM} (\partial \mathcal{S}/\partial T_{SM})_P$ is the modified heat capacity. The inversion curve, separating the cooling ($\mu > 0$) and heating ($\mu < 0$) phases, occurs when $\mu = 0$, yielding the condition:
\begin{equation}
    T_i = V \left( \frac{\partial T_{SM}}{\partial V} \right)_P = \frac{r_+}{3} \left( \frac{\partial T_{SM}}{\partial r_+} \right)_P.
    \label{Ti_condition}
\end{equation}
To isolate the inversion pressure $P_i$, we separate $T_0(r_+, P)$ into pressure-dependent and independent parts: $T_0 = 2Pr_+ + A(r_+)$, where $A(r_+)$ collects the remaining terms. Substituting $T_{SM} = [2Pr_+ + A(r_+)]f(r_+)$ into Eq. (\ref{Ti_condition}) allows us to solve exactly for the inversion pressure:
\begin{equation}
    P_{i} = \frac{r_+ A'(r_+) f(r_+) - A(r_+) [3 f(r_+) - r_+ f'(r_+)]}{2r_+ [2f(r_+) - r_+ f'(r_+)]},
    \label{Pi_SM}
\end{equation}
where primes denote derivatives with respect to $r_+$. The corresponding inversion temperature is found by substituting $P_i$ back into $T_{SM}$. Eq. (\ref{Pi_SM}) dictates that the inversion phase boundary is governed by a delicate interplay between the ModMax field parameter $\gamma$, massive graviton interactions, and the non-extensive SM parameters $R$ and $\delta$.

\section{Results and Discussion}
\label{sec4}

To thoroughly understand the throttling process and the local stability of the black hole, we numerically analyse the specific heat, the Joule-Thomson coefficient, the isenthalpic curves, and the inversion phase boundaries. We consider two distinct regimes of the dRGT-like massive gravity parameters to observe the robustness of our results:
\begin{itemize}
    \item \textbf{Case A (Weak Massive Gravity):} $m_g=0.4, \Lambda=-0.5, C=0.1, c_1=1, c_2=0.2$.
    \item \textbf{Case B (Strong Massive Gravity):} $m_g=0.8, \Lambda=-0.5, C=0.4, c_1=-5, c_2=7$.
\end{itemize}
For both cases, we set the electric charge $q=1$. If not mentioned, $\gamma = 0.5$
$\delta = 0.1$ and
$R = 0.05$. We systematically vary the ModMax parameter ($\gamma$) and the Sharma-Mittal non-extensive parameters ($\delta, R$) to isolate their physical impacts.

\subsection{Specific Heat and Local Stability}

\begin{figure*}[htbp]
    \centering
    \subfloat[Case A: Varying $\gamma$]{\includegraphics[width=0.42\linewidth]{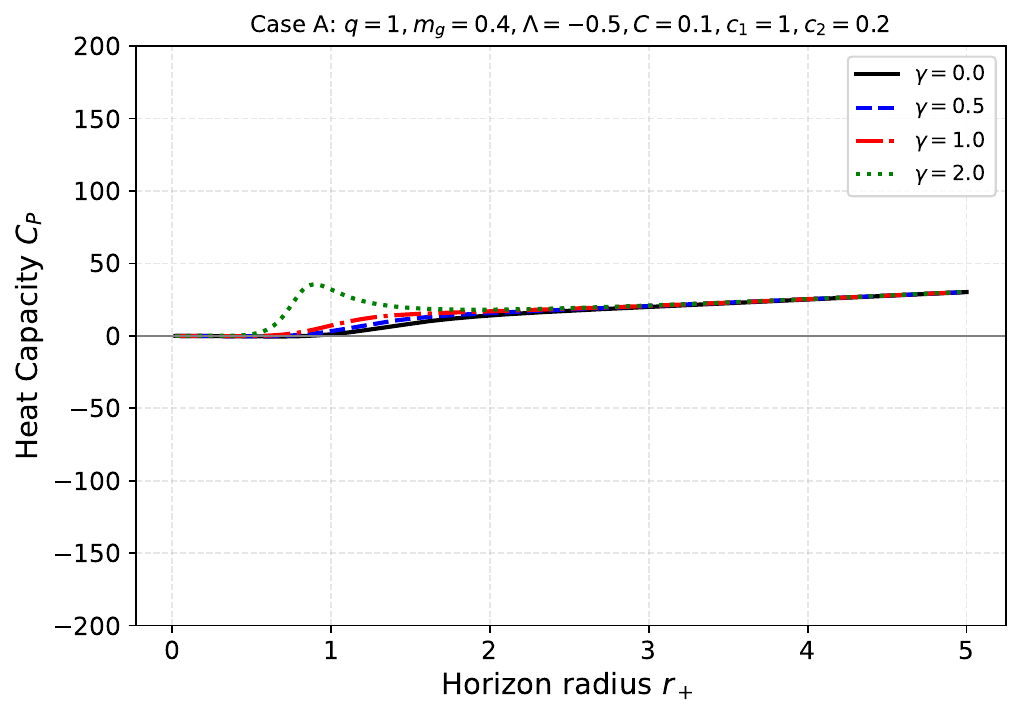}}
    \hfill
    \subfloat[Case B: Varying $\gamma$]{\includegraphics[width=0.42\linewidth]{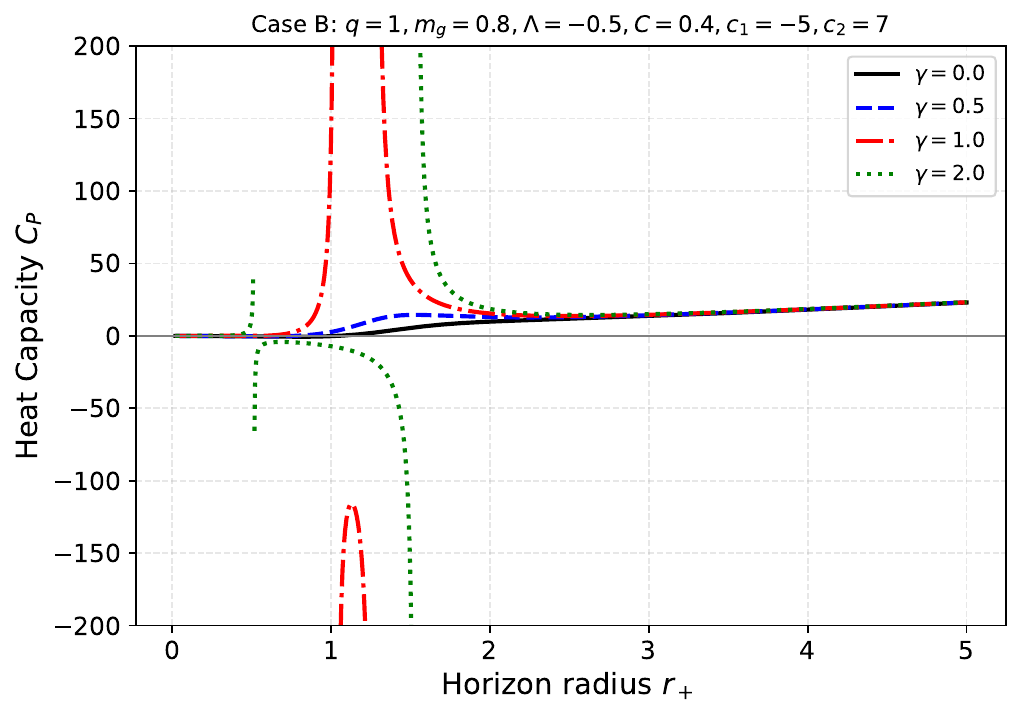}}
    \\
    \subfloat[Case A: Varying $\delta$]{\includegraphics[width=0.42\linewidth]{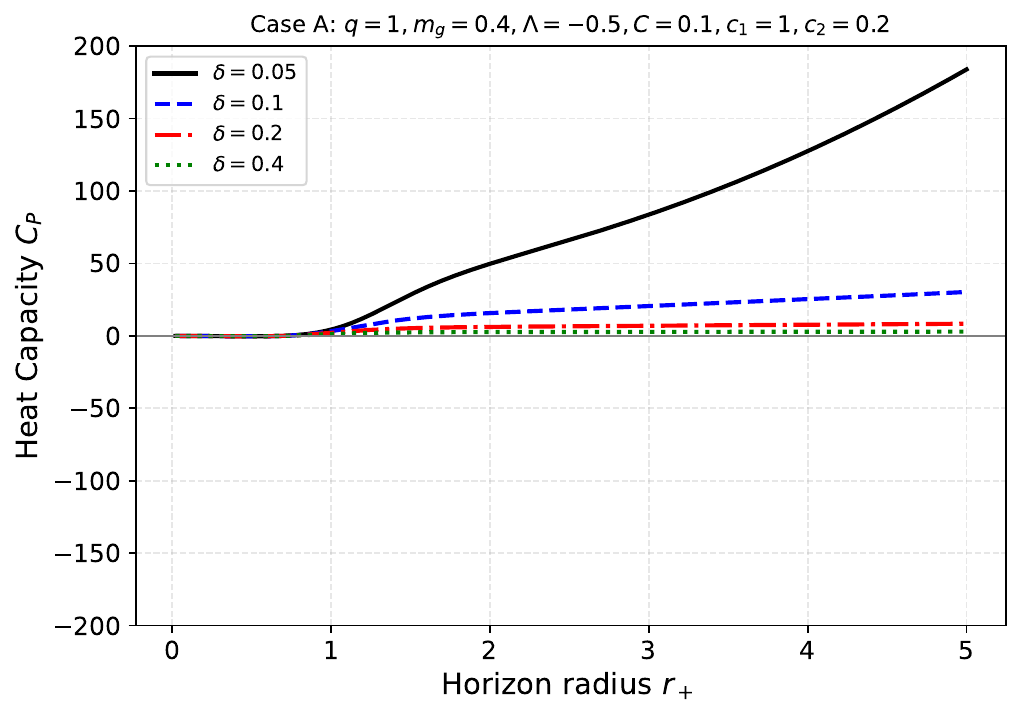}}
    \hfill
    \subfloat[Case B: Varying $\delta$]{\includegraphics[width=0.42\linewidth]{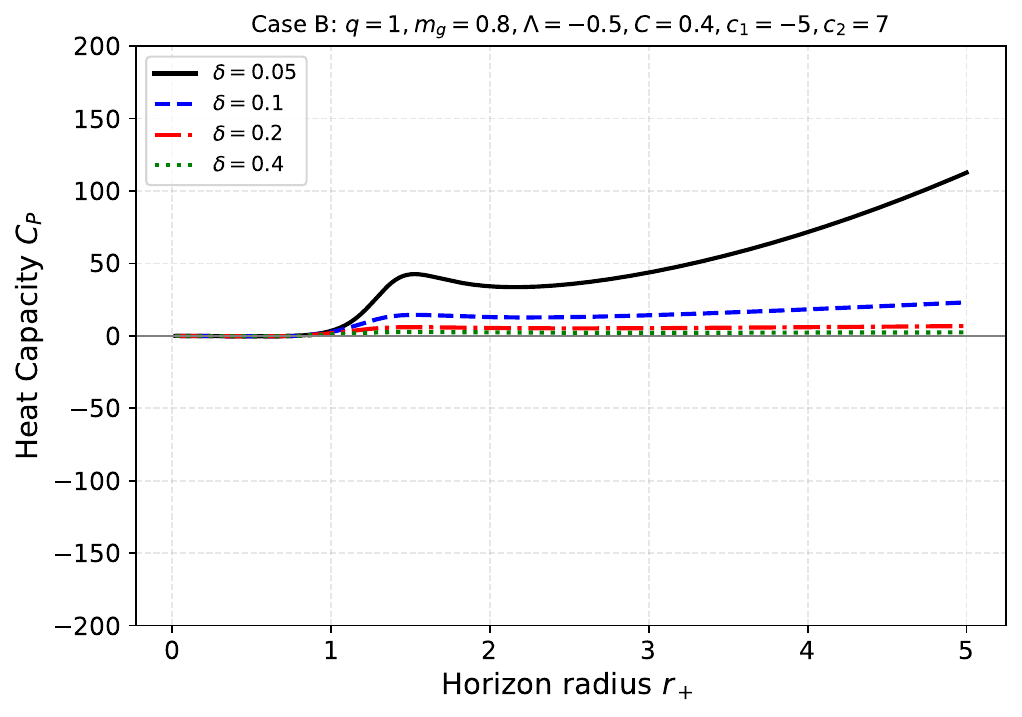}}
    \\
    \subfloat[Case A: Varying $R$]{\includegraphics[width=0.42\linewidth]{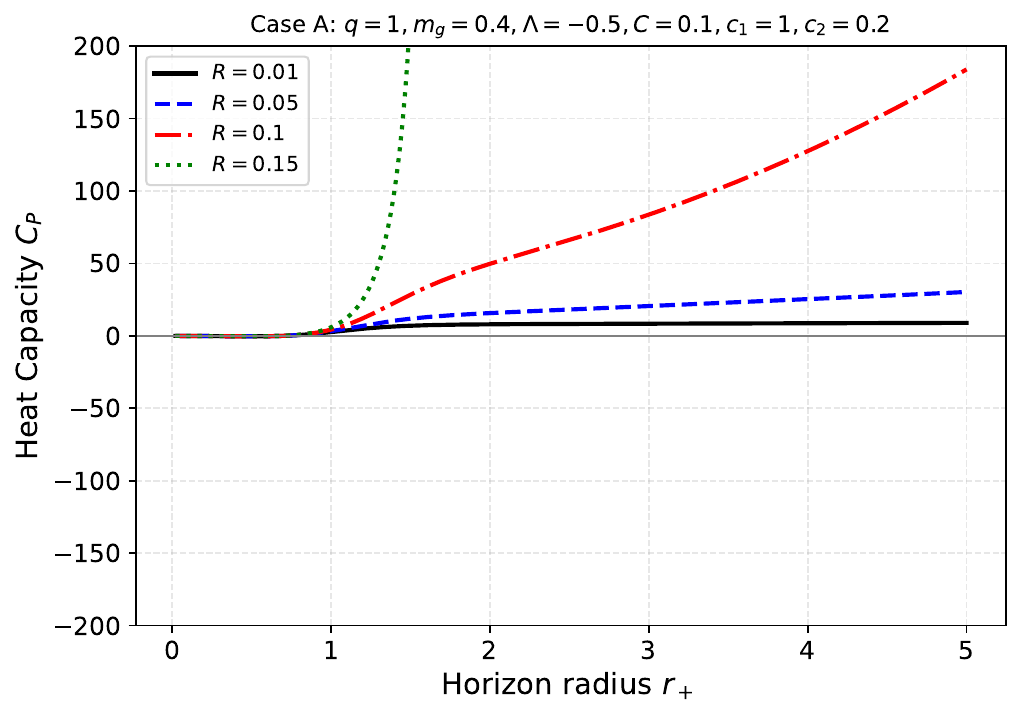}}
    \hfill
    \subfloat[Case B: Varying $R$]{\includegraphics[width=0.42\linewidth]{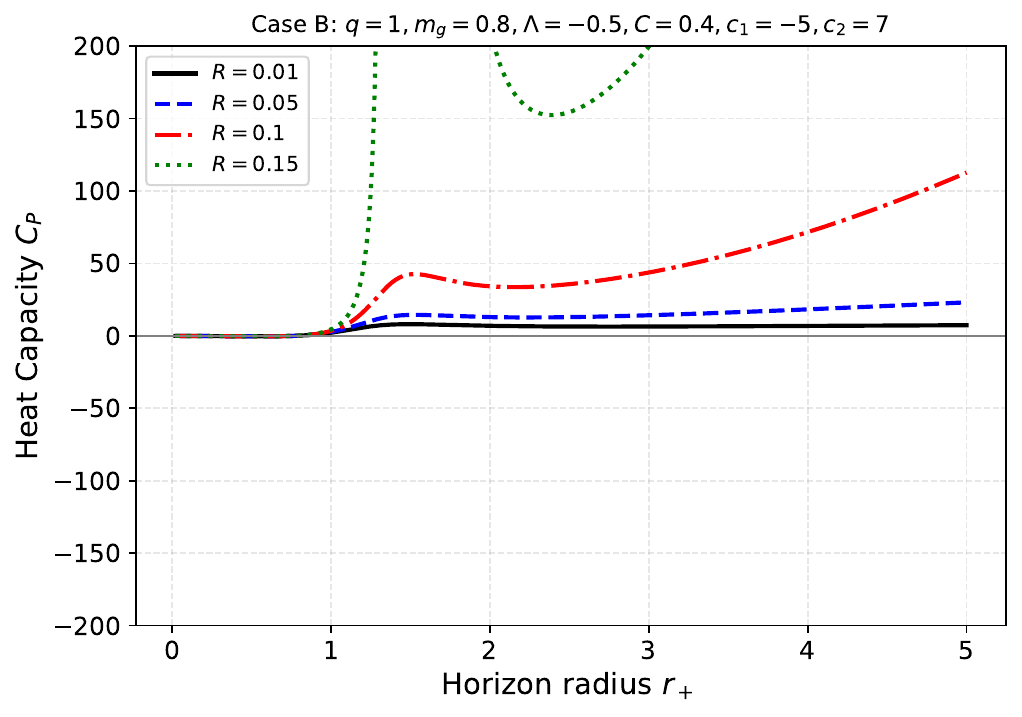}}
    \caption{Specific heat $C_P$ versus horizon radius $r_+$ under varying ModMax and Sharma-Mittal parameters for weak (Case A) and strong (Case B) massive gravity regimes.}
    \label{fig:Cp}
\end{figure*}

The local thermodynamic stability of the black hole is governed by the sign of the heat capacity $C_P$ at constant pressure. The behaviour of $C_P$ is illustrated in Fig. \ref{fig:Cp}.

Our analysis reveals that the ModMax parameter $\gamma$ predominantly affects the near-horizon (small $r_+$) regime. In the weak massive gravity regime (Case A), increasing $\gamma$ introduces a distinct local maximum at small radii. For strong massive gravity (Case B), the specific heat exhibits high volatility and multiple divergences at higher values of $\gamma$. However, as $\gamma$ decreases, the nonlinear conformal effects smooth out the specific heat profile, eliminating the unstable intermediate regions and stabilising the black hole.

The non-extensive SM parameters, $\delta$ and $R$, exhibit competing macroscopic effects. Increasing $\delta$ strongly suppresses the growth of the specific heat at larger horizon radii, implying that higher-order microstate correlations effectively constrain the thermal fluctuations of large black holes. Conversely, increasing $R$ amplifies $C_P$ at larger radii, eventually driving the system toward massive divergences in the strong massive gravity regime.

\subsection{Joule-Thomson Coefficient}

\begin{figure*}[htbp]
    \centering
    \subfloat[Case A: Varying $\gamma$]{\includegraphics[width=0.43\linewidth]{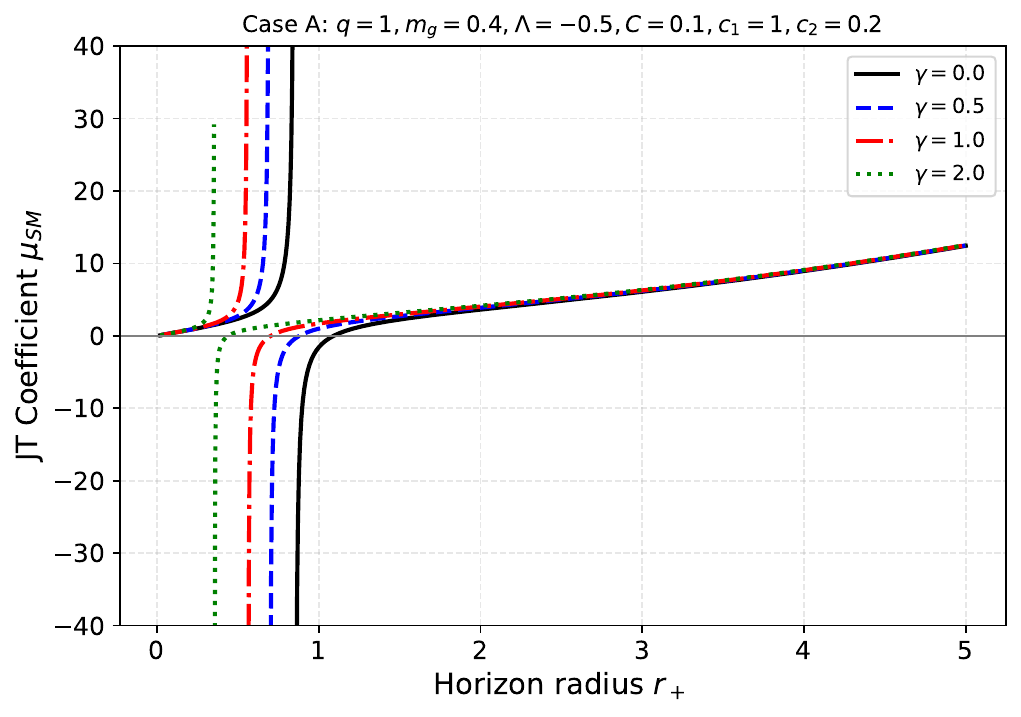}}
    \hfill
    \subfloat[Case B: Varying $\gamma$]{\includegraphics[width=0.43\linewidth]{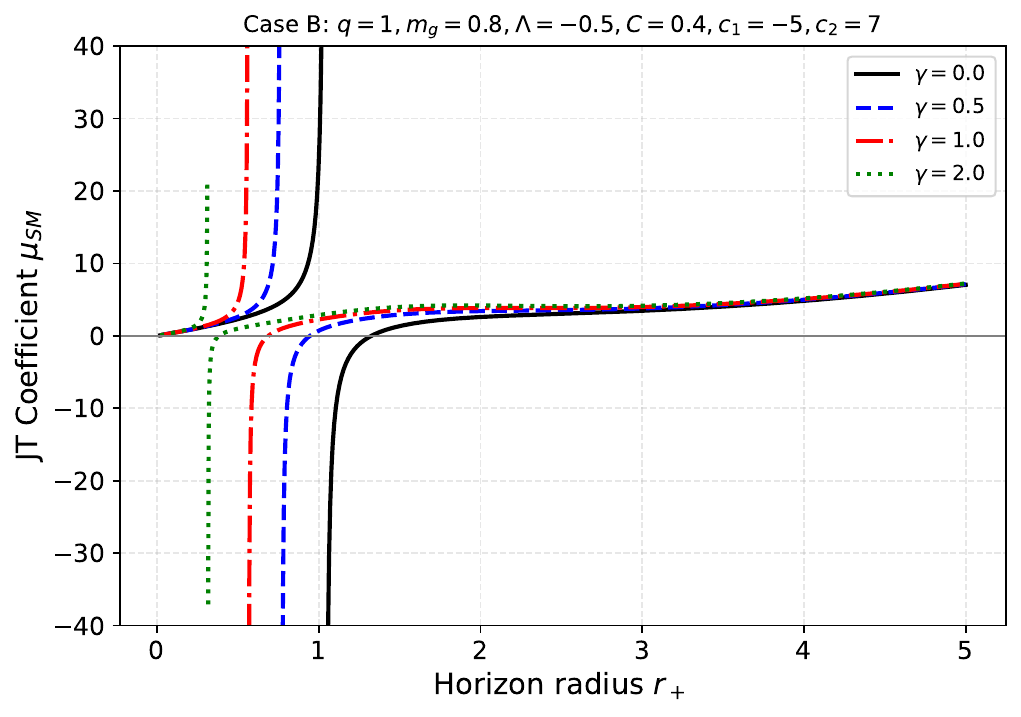}}
    \\
    \subfloat[Case A: Varying $\delta$]{\includegraphics[width=0.43\linewidth]{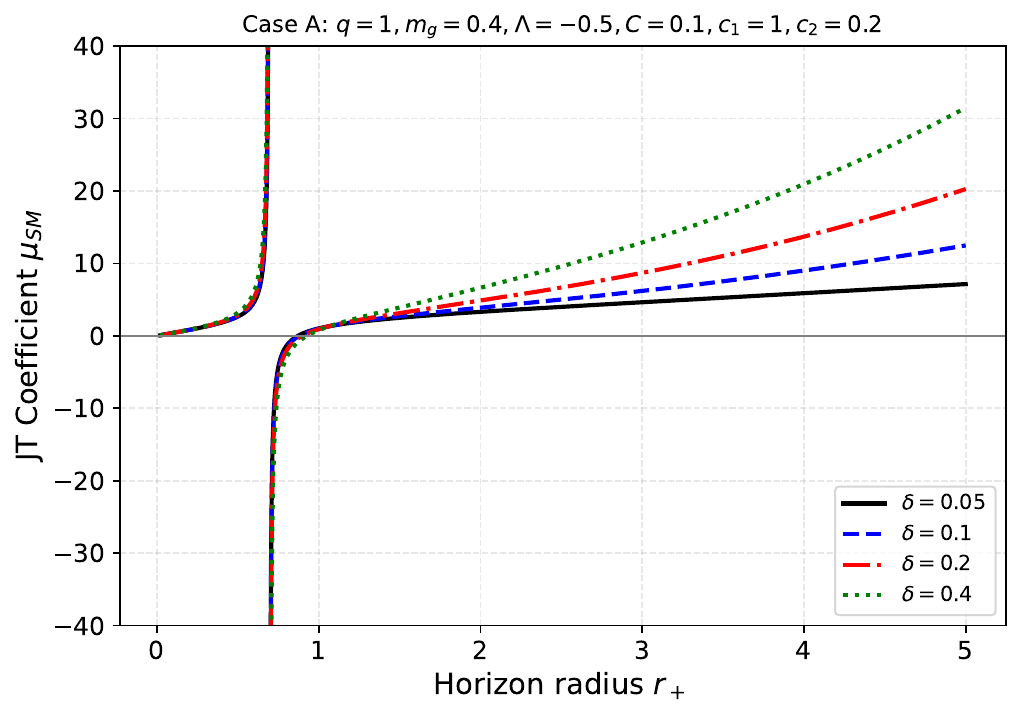}}
    \hfill
    \subfloat[Case B: Varying $\delta$]{\includegraphics[width=0.43\linewidth]{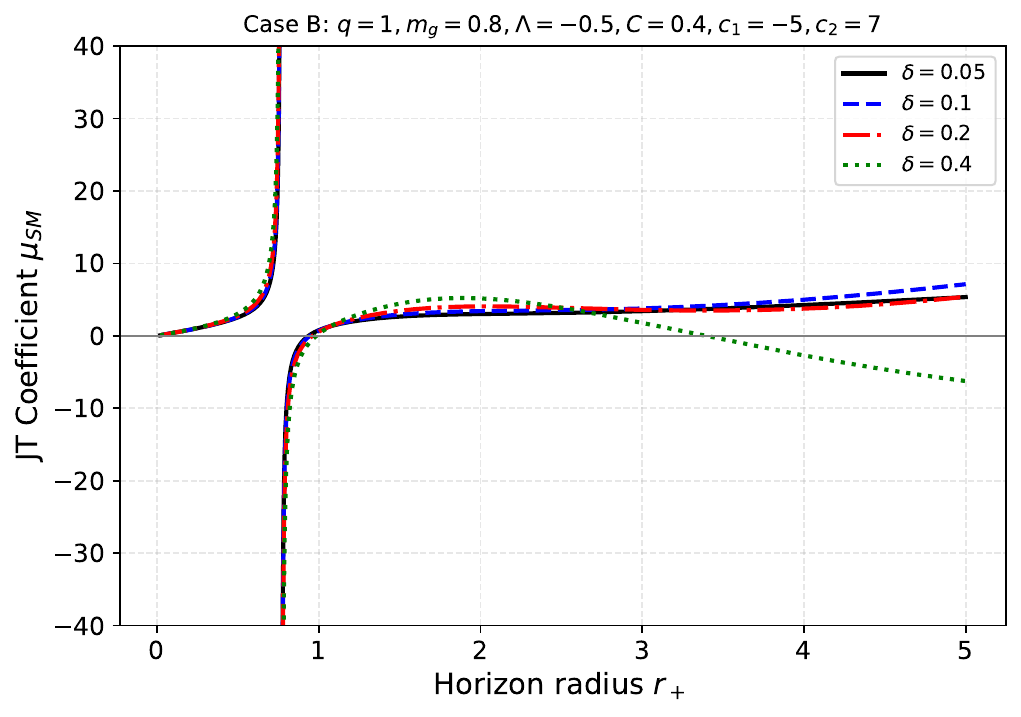}}
\\
    \subfloat[Case A: Varying $R$]{\includegraphics[width=0.43\linewidth]{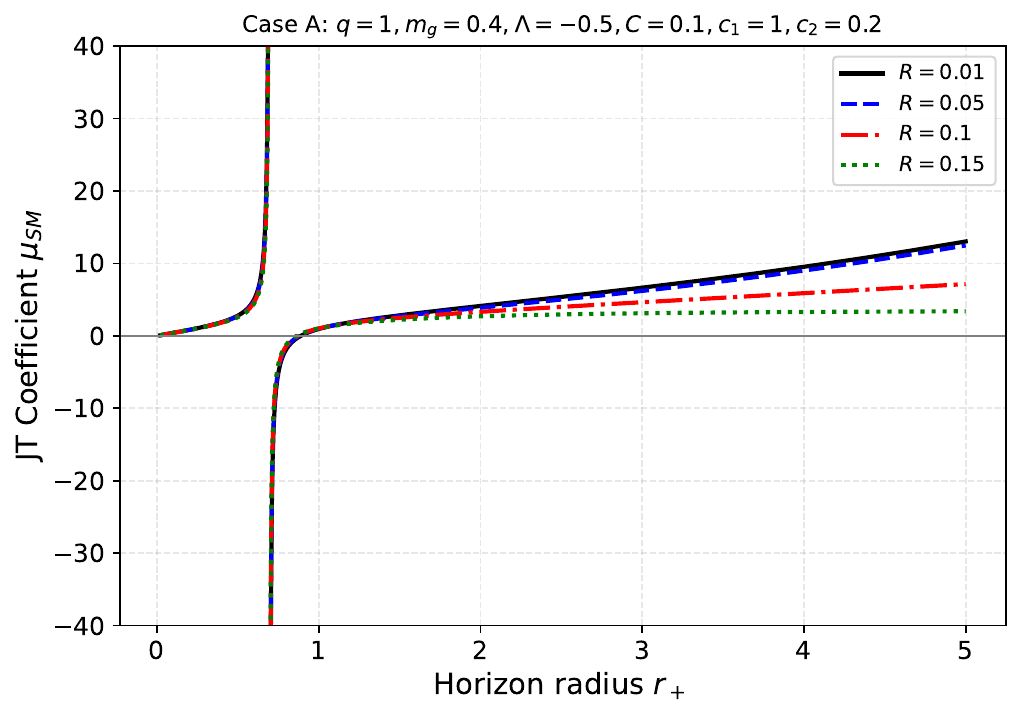}}
    \hfill
    \subfloat[Case B: Varying $R$]{\includegraphics[width=0.43\linewidth]{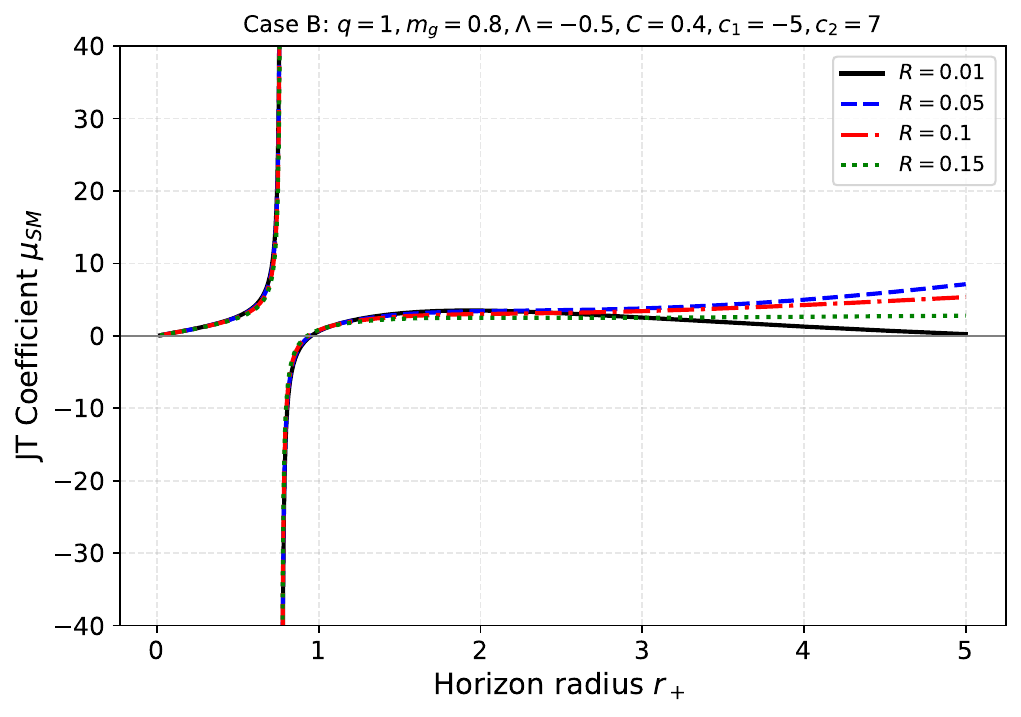}}
    
    \caption{Joule-Thomson coefficient $\mu_{SM}$ versus horizon radius $r_+$. The divergence points correspond to the inversion radii of the black holes.}
    \label{fig:JT}
\end{figure*}

The Joule-Thomson coefficient $\mu_{SM}$ determines the heating and cooling phases of the black hole during an isenthalpic expansion. The boundary between these phases is marked by the divergence of $\mu_{SM}$, which geometrically corresponds to the inversion radius. This behaviour is depicted in Fig. \ref{fig:JT}.

Increasing the ModMax parameter systematically shifts the divergence point to smaller horizon radii in both Case A and Case B. This indicates that the transition radius is highly sensitive to the conformal nonlinearities of the electromagnetic field; a weaker effective electromagnetic repulsion (due to large $\gamma$) causes the throttling transition to occur at smaller black hole sizes. Furthermore, in the large $r_+$ limit, increasing the non-extensive parameter $\delta$ raises the value of $\mu_{SM}$ for case A, whereas an opposite scenario occurs in case B. However, $R$ seems to have an opposite impact on  $\mu_{SM}$ in comparison to that of $\delta$.

\subsection{Isenthalpic Curves and Inversion Phase Boundary}

\begin{figure*}[htbp]
    \centering
    \subfloat[Case A: Varying $\gamma$]{\includegraphics[width=0.32\linewidth]{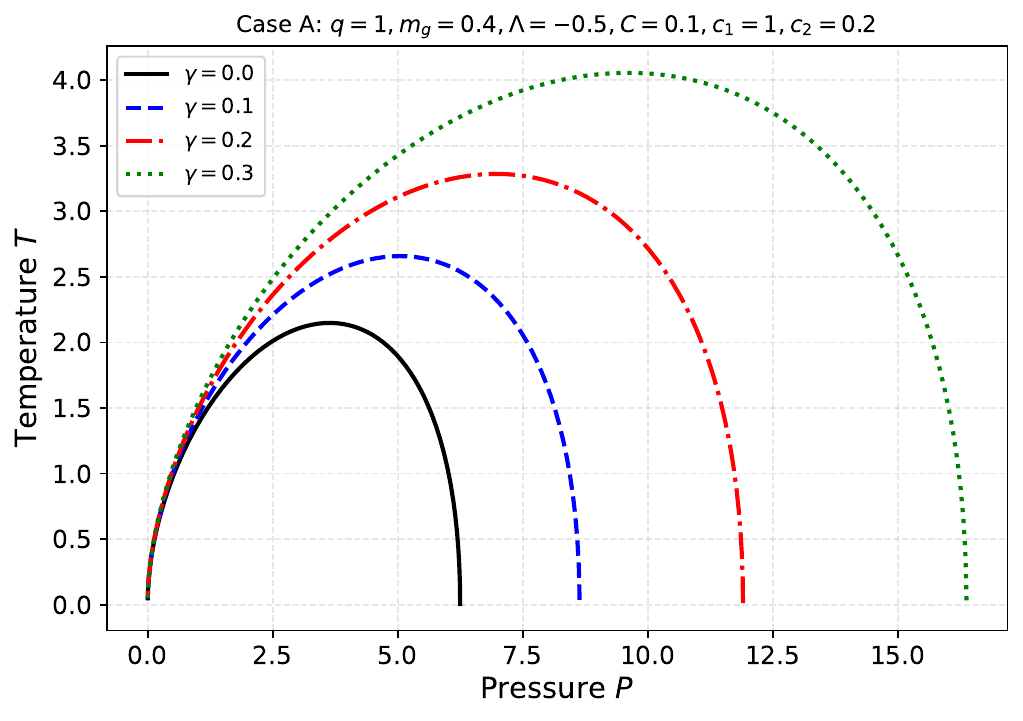}}
    \hfill
    \subfloat[Case A: Varying $\delta$]{\includegraphics[width=0.32\linewidth]{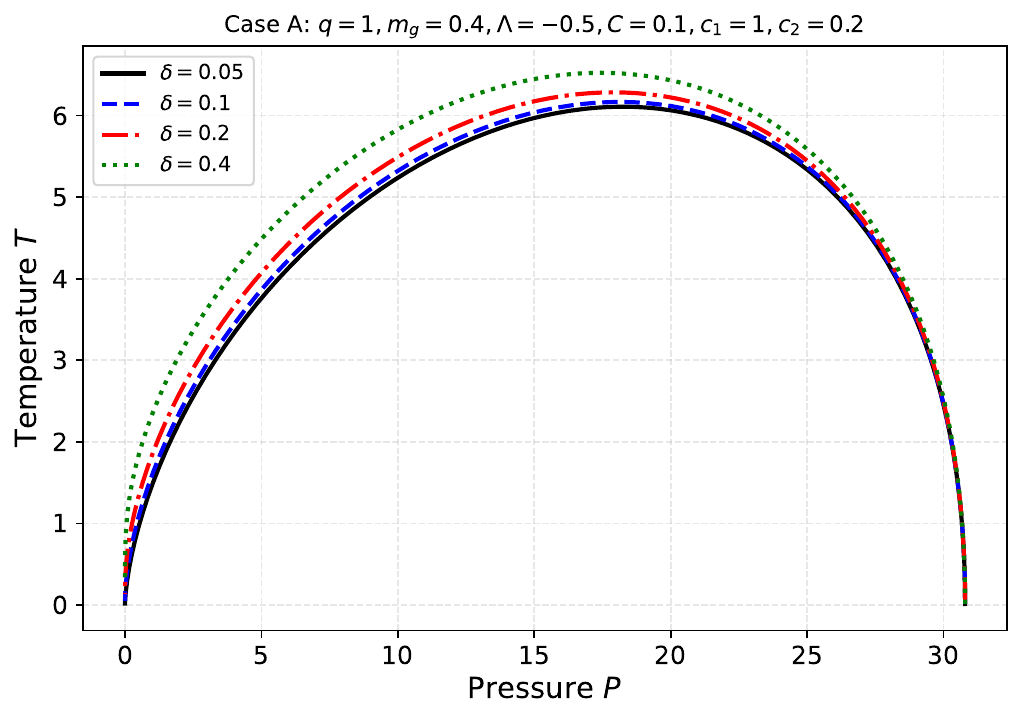}}
    \hfill
    \subfloat[Case A: Varying $R$]{\includegraphics[width=0.32\linewidth]{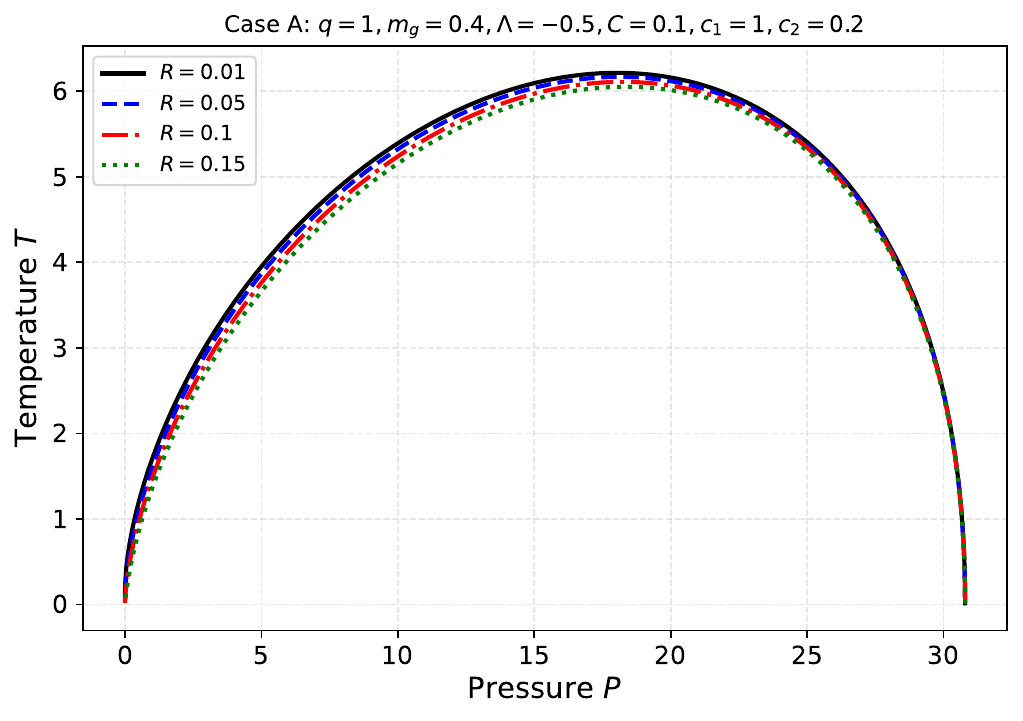}}
    \\
    \subfloat[Case B: Varying $\gamma$]{\includegraphics[width=0.32\linewidth]{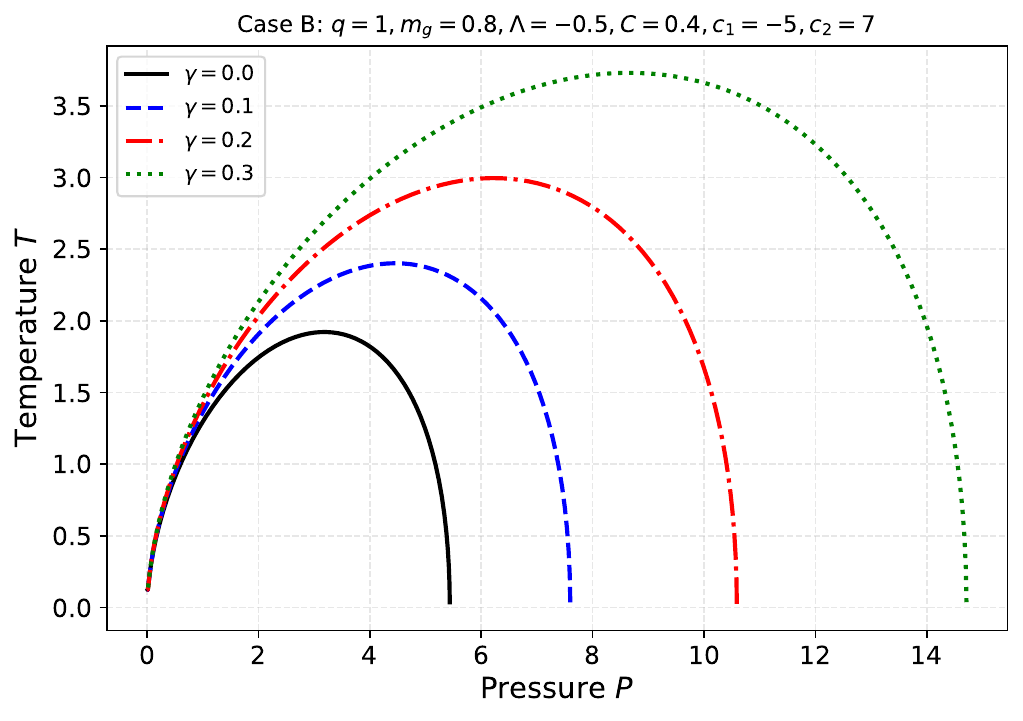}}
    \hfill
    \subfloat[Case B: Varying $\delta$]{\includegraphics[width=0.32\linewidth]{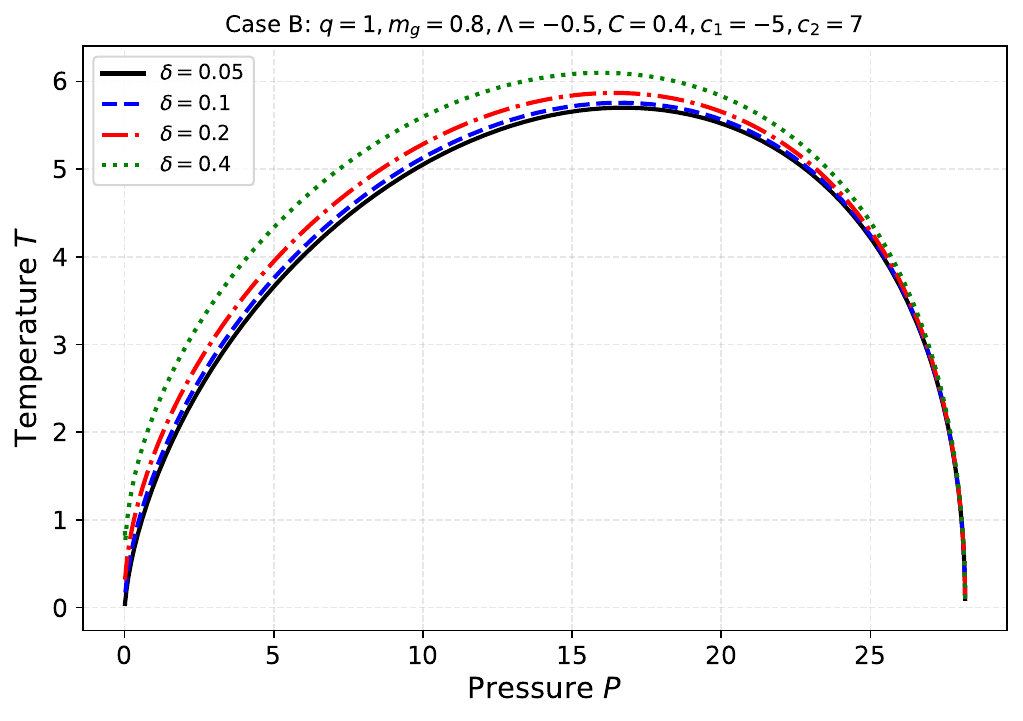}}
    \hfill
    \subfloat[Case B: Varying $R$]{\includegraphics[width=0.32\linewidth]{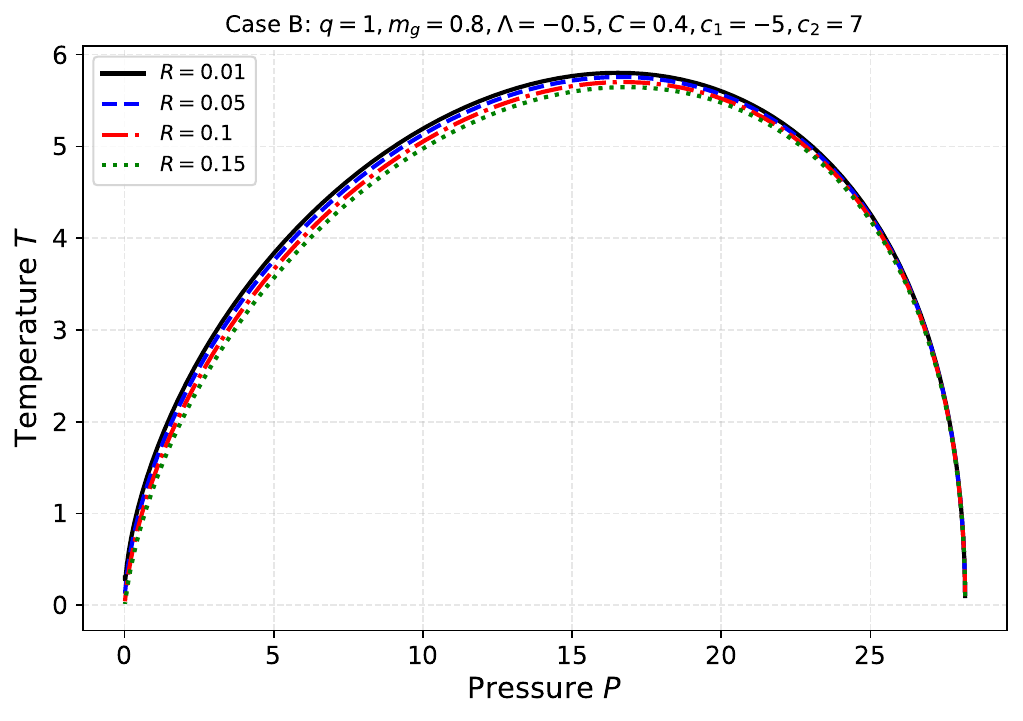}}
    \caption{Isenthalpic curves in the $T-P$ plane under varying ModMax and Sharma-Mittal parameters for weak (Case A) and strong (Case B) massive gravity regimes.}
    \label{fig:Iso}
\end{figure*}

\begin{figure*}[htbp]
    \centering
    \subfloat[Case A: Varying $\gamma$]{\includegraphics[width=0.32\linewidth]{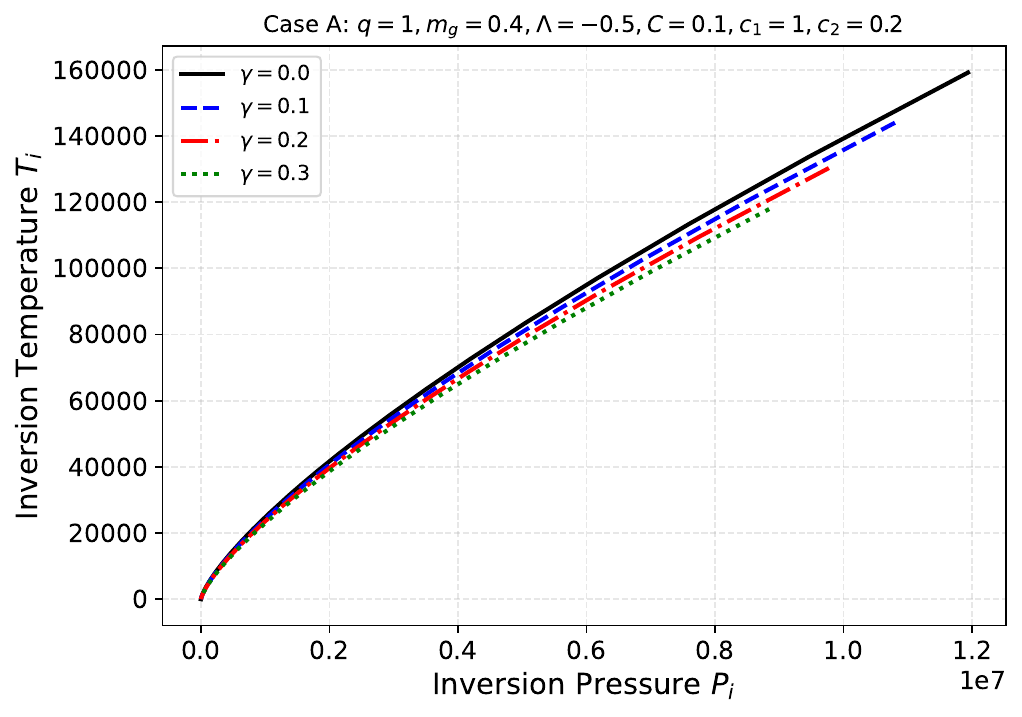}}
    \hfill
    \subfloat[Case A: Varying $\delta$]{\includegraphics[width=0.32\linewidth]{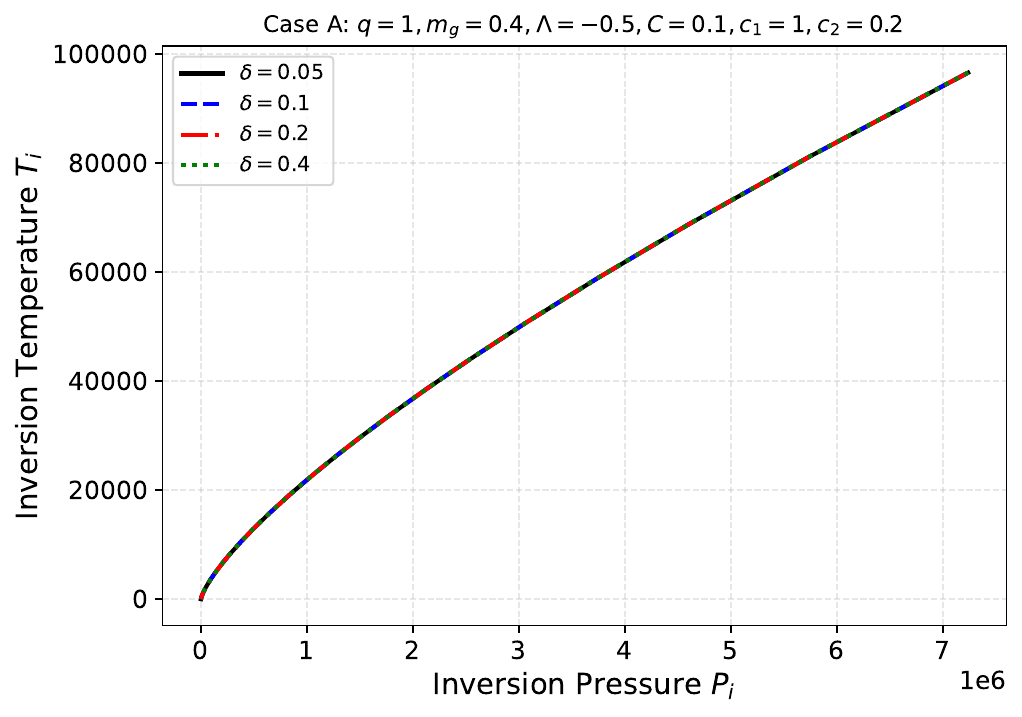}}
    \hfill
    \subfloat[Case A: Varying $R$]{\includegraphics[width=0.32\linewidth]{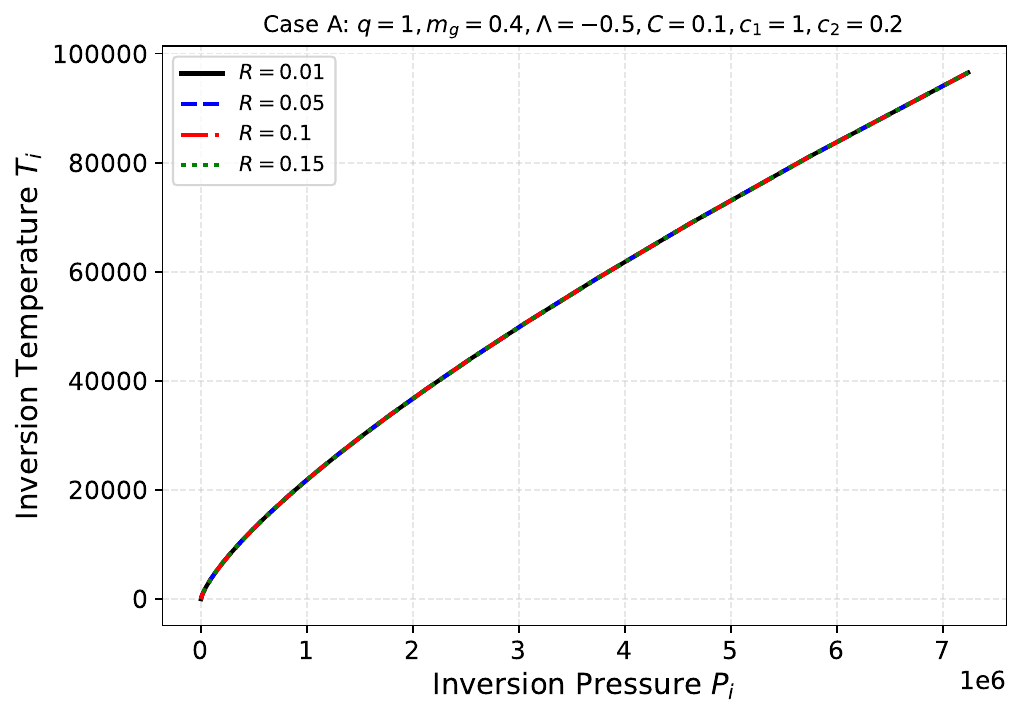}}
    \\
    \subfloat[Case B: Varying $\gamma$]{\includegraphics[width=0.32\linewidth]{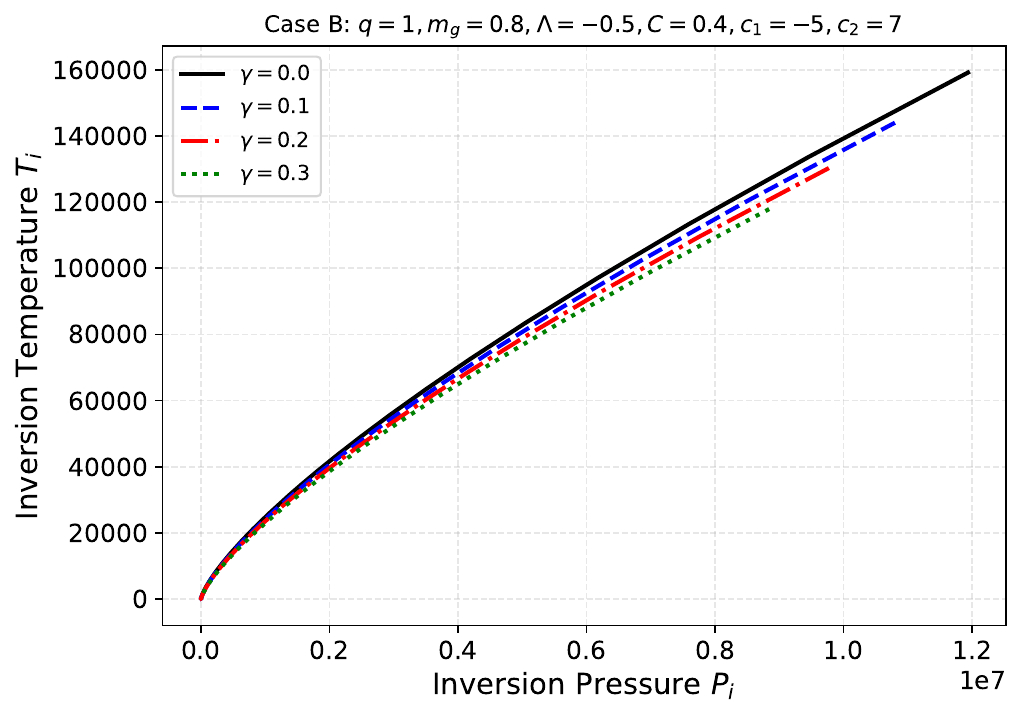}}
    \hfill
    \subfloat[Case B: Varying $\delta$]{\includegraphics[width=0.32\linewidth]{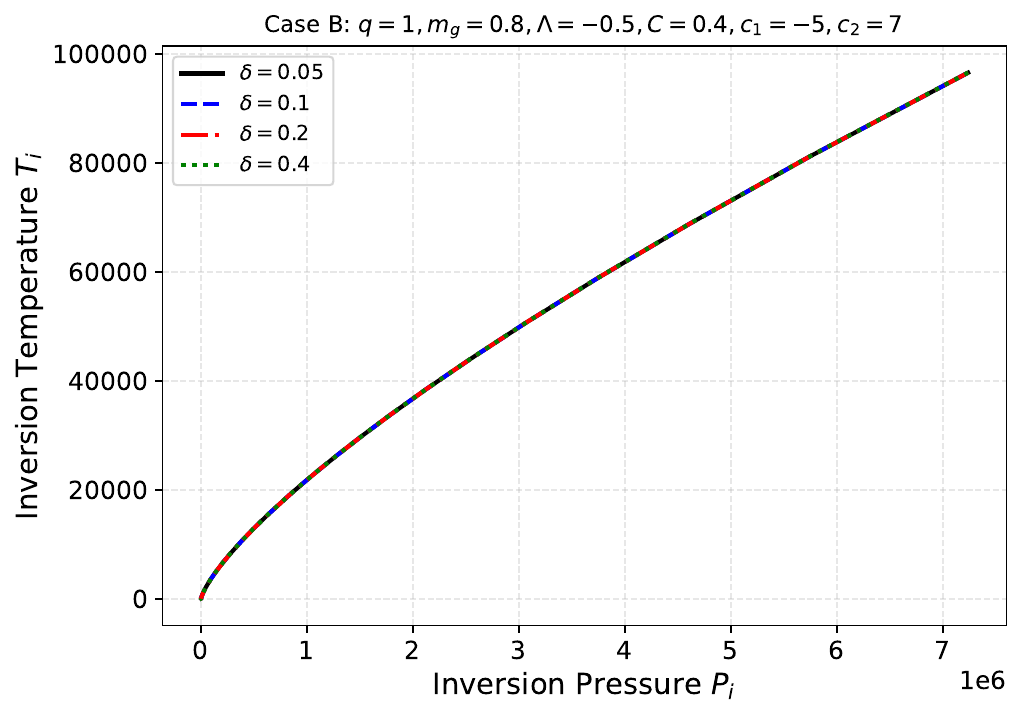}}
    \hfill
    \subfloat[Case B: Varying $R$]{\includegraphics[width=0.32\linewidth]{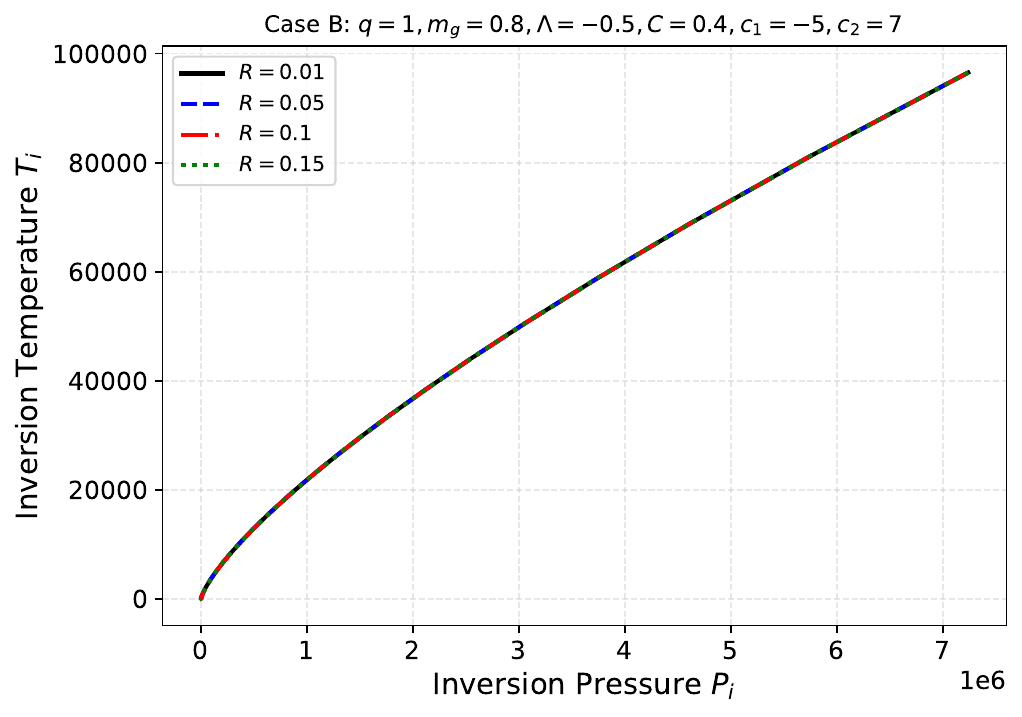}}
    \caption{Inversion curves in the $T-P$ plane representing the boundary between heating and cooling phases.}
    \label{fig:Inv}
\end{figure*}

\begin{figure*}[htbp]
    \centering
    \subfloat[Case A: Combined Iso/Inv]{\includegraphics[width=0.45\linewidth]{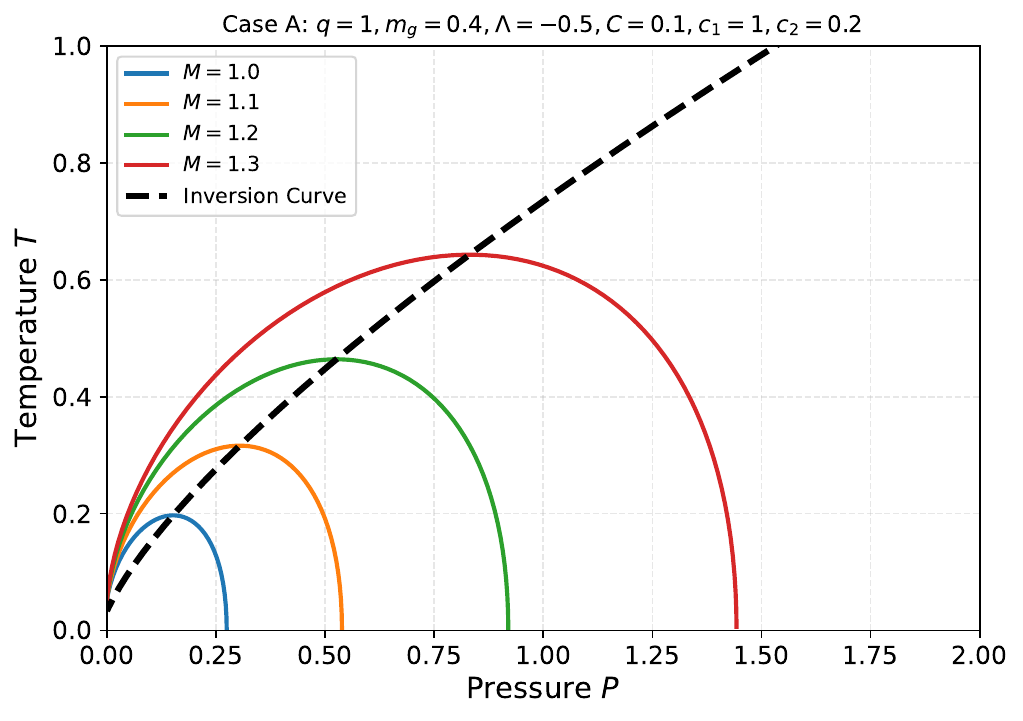}}
    \hfill
    \subfloat[Case B: Combined Iso/Inv]{\includegraphics[width=0.45\linewidth]{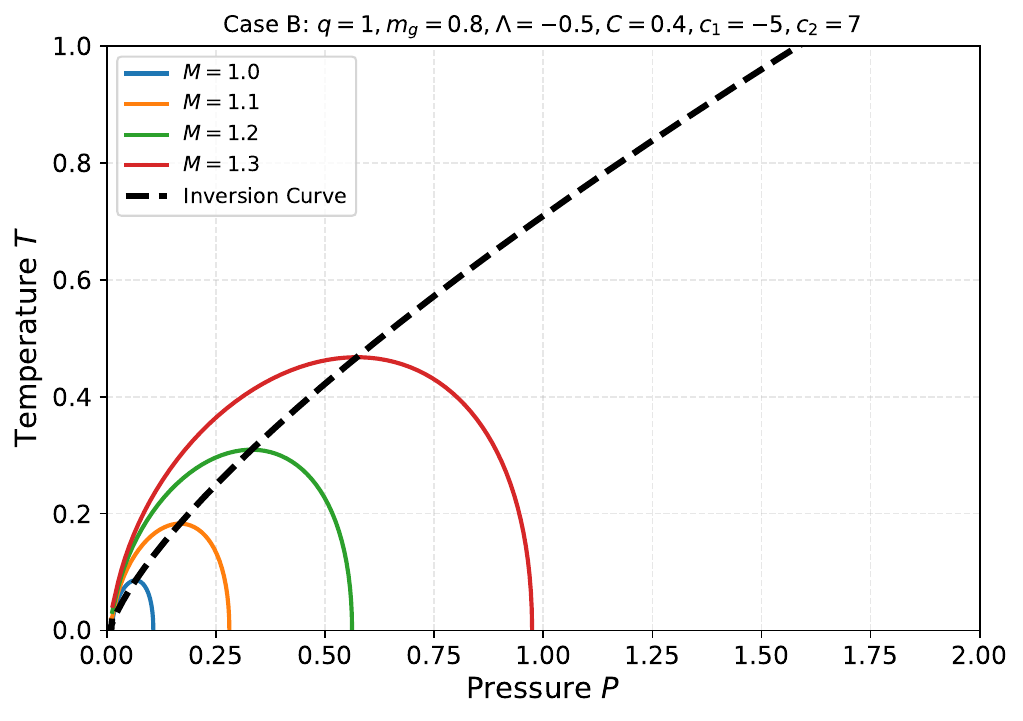}}
    \caption{Combined isenthalpic and inversion curves in the $T-P$ plane. The inversion curve strictly passes through the exact maximum of each isenthalpic curve.}
    \label{fig:CombinedIsoInv}
\end{figure*}

To construct the global phase space of the JT expansion, we plot the isenthalpic curves alongside the inversion curves in the $T-P$ plane. The isolated isenthalpic and inversion curves for varying parameters are presented in Figs. \ref{fig:Iso} and \ref{fig:Inv}, respectively, while their combined behavior is shown in Fig. \ref{fig:CombinedIsoInv}.

As observed in Fig. \ref{fig:CombinedIsoInv}, the isenthalpic curves form distinct parabolic arcs. The region with a positive slope ($\mu_{SM} > 0$) represents the cooling phase, and the region with a negative slope ($\mu_{SM} < 0$) represents the heating phase. The inversion curve strictly passes through the exact peak of each isenthalpic arc. 

The impact of the ModMax parameter is massive. As shown in Fig. \ref{fig:Iso}, increasing $\gamma$ drastically expands the isenthalpic curves, pushing the peak inversion temperatures and pressures to significantly higher values. However, looking at the global inversion curve ($T_i$ vs $P_i$) in Fig. \ref{fig:Inv}, increasing $\gamma$ significantly shifts the inversion boundary downward. Because the cooling region lies above the inversion curve in the standard AdS formalism, a lower curve implies that the nonlinear conformal effects of the ModMax field effectively increase the available cooling domain for the black hole at any given pressure.

Interestingly, while $\delta$ and $R$ profoundly impact the local specific heat and the exact location of the divergences in the JT coefficient, their effect on the global $T_i$ vs $P_i$ inversion boundary is virtually negligible at the macroscopic scale. The inversion curves for varying $\delta$ and $R$ (middle and right panels of Fig. \ref{fig:Inv}) overlap almost entirely. This demonstrates that non-extensive Sharma-Mittal entropy primarily governs the local microstate correlations and phase stability, but the global heating-cooling transition boundary is overwhelmingly dictated by the macroscopic spacetime geometry (massive gravity) and the background electromagnetic field (ModMax). Consistent with other AdS black hole models, the inversion curve consists of a single, monotonically increasing branch, lacking the upper inversion curve typically seen in classical van der Waals fluids.

\section{$P$--$V$ Criticality and Phase Structure}
\label{sec:pv-criticality}

The equation of state of the Sharma--Mittal corrected ModMax-dRGT black hole can be written as \cite{johnson2014holographic,kubizvnak2012p}
\begin{equation}
	P=
	\frac{T}{2r_+f(r_+)}
	-\frac{1}{8\pi r_+^2}
	+\frac{q^2e^{-\gamma}}{8\pi r_+^4}
	-\frac{m_g^2Cc_1}{8\pi r_+}
	-\frac{m_g^2C^2c_2}{8\pi r_+^2},
	\label{eq:eos_criticality}
\end{equation}
where
\begin{equation}
	f(r_+)=
	\left(1+\delta\pi r_+^2\right)^\alpha,
	\qquad
	\alpha=1-\frac{R}{\delta}.
	\label{eq:sm_factor_criticality}
\end{equation}
The critical point corresponds to the inflection point of the isotherm in the
$P-r_+$ plane and is determined by
\begin{equation}
	\left(\frac{\partial P}{\partial r_+}\right)_T=0,
	\qquad
	\left(\frac{\partial^2P}{\partial r_+^2}\right)_T=0.
	\label{eq:critical_conditions}
\end{equation}
Differentiating Eq.~\eqref{eq:eos_criticality}, one obtains
\begin{equation}
	\left(\frac{\partial P}{\partial r_+}\right)_T
	=
	-\frac{T}{2}
	\left[
	\frac{1}{r_+^2f}
	+
	\frac{2\alpha\delta\pi}
	{f(1+\delta\pi r_+^2)}
	\right]
	+\frac{1}{4\pi r_+^3}
	-\frac{q^2e^{-\gamma}}{2\pi r_+^5}
	+\frac{m_g^2Cc_1}{8\pi r_+^2}
	+\frac{m_g^2C^2c_2}{4\pi r_+^3}.
	\label{eq:first_derivative_p}
\end{equation}
From the first condition in Eq.~\eqref{eq:critical_conditions}, the critical temperature follows as
\begin{equation}
	T_c=
	\frac{
		\frac{1}{2\pi r_c^3}
		-\frac{q^2e^{-\gamma}}{\pi r_c^5}
		+\frac{m_g^2Cc_1}{4\pi r_c^2}
		+\frac{m_g^2C^2c_2}{2\pi r_c^3}
	}{
		\frac{1}{r_c^2f_c}
		+
		\frac{2\alpha\delta\pi}
		{f_c(1+\delta\pi r_c^2)}
	},
	\label{eq:critical_temperature}
\end{equation}
where $f_c=f(r_c)$.

The second derivative condition gives a single algebraic equation for the critical radius,
\begin{equation}
	\frac{T_c}{2}\,\mathcal{F}(r_c)
	-\frac{3}{4\pi r_c^4}
	+\frac{5q^2e^{-\gamma}}{2\pi r_c^6}
	-\frac{m_g^2Cc_1}{4\pi r_c^3}
	-\frac{3m_g^2C^2c_2}{4\pi r_c^4}
	=0,
	\label{eq:critical_radius_equation}
\end{equation}
with
\begin{equation}
	\mathcal{F}(r)
	=
	\frac{2}{r^3f}
	+
	\frac{4\alpha\delta\pi}
	{rf(1+\delta\pi r^2)}
	+
	\frac{4\alpha(\alpha+1)\delta^2\pi^2r}
	{f(1+\delta\pi r^2)^2}.
	\label{eq:F_factor_criticality}
\end{equation}
Once $r_c$ is obtained, the critical pressure is determined from the equation of state,
\begin{equation}
	P_c=
	\frac{T_c}{2r_cf_c}
	-\frac{1}{8\pi r_c^2}
	+\frac{q^2e^{-\gamma}}{8\pi r_c^4}
	-\frac{m_g^2Cc_1}{8\pi r_c}
	-\frac{m_g^2C^2c_2}{8\pi r_c^2}.
	\label{eq:critical_pressure}
\end{equation}

The existence of a solution $(r_c,T_c,P_c)$ demonstrates the presence of a van der Waals-like critical point associated with a small/large black-hole phase transition. The ModMax parameter enters through the effective charge combination
\begin{equation}
	q_{\rm eff}^{\,2}=q^2e^{-\gamma},
	\label{eq:qeff}
\end{equation}
which suppresses the electromagnetic contribution and shifts the critical point toward smaller horizon radii. In contrast, the Sharma--Mittal parameters $(R,\delta)$ modify the thermal sector through the factor $f(r_+)$, while the massive-gravity couplings directly affect the geometric structure of the equation of state and therefore play a dominant role in determining the critical behaviour.

\section{Gibbs Free Energy and Global Thermodynamic Stability}
\label{sec:gibbs-free-energy}

To investigate the global thermodynamic stability of the system, we introduce the Gibbs free energy \cite{york1986black}
\begin{equation}
	G=M-T_{\rm SM}S_{\rm SM}.
	\label{eq:gibbs_definition}
\end{equation}
Here $M$ is the black-hole enthalpy,
\begin{equation}
	M=
	\frac{r_+}{2}
	+\frac{4\pi Pr_+^3}{3}
	+\frac{q^2e^{-\gamma}}{2r_+}
	+\frac{m_g^2Cr_+}{2}
	\left(
	\frac{c_1r_+}{2}
	+c_2C
	\right),
	\label{eq:enthalpy_gibbs}
\end{equation}
and the Sharma--Mittal entropy is
\begin{equation}
	S_{\rm SM}
	=
	\frac{1}{R}
	\left[
	\left(1+\delta\pi r_+^2\right)^{R/\delta}
	-1
	\right].
	\label{eq:sm_entropy_gibbs}
\end{equation}
The modified Hawking temperature takes the form
\begin{equation}
	T_{\rm SM}
	=
	T_0
	\left(1+\delta\pi r_+^2\right)^{1-R/\delta},
	\label{eq:sm_temperature_gibbs}
\end{equation}
where
\begin{equation}
	T_0=
	2Pr_+
	+
	\frac{1}{4\pi}
	\left(
	\frac{1}{r_+}
	-
	\frac{q^2e^{-\gamma}}{r_+^3}
	\right)
	+
	\frac{m_g^2C}{4\pi}
	\left(
	c_1+\frac{c_2C}{r_+}
	\right).
	\label{eq:T0_gibbs}
\end{equation}
Substituting Eqs.~\eqref{eq:enthalpy_gibbs}--\eqref{eq:sm_temperature_gibbs} into Eq.~\eqref{eq:gibbs_definition}, the Gibbs free energy is obtained explicitly as
\begin{align}
	G(r_+,P)
	=&
	\frac{r_+}{2}
	+\frac{4\pi Pr_+^3}{3}
	+\frac{q^2e^{-\gamma}}{2r_+}
	+\frac{m_g^2Cr_+}{2}
	\left(
	\frac{c_1r_+}{2}
	+c_2C
	\right)
	\nonumber\\
	&-
	\frac{1}{R}
	\left[
	\left(1+\delta\pi r_+^2\right)^{R/\delta}
	-1
	\right]
	T_0
	\left(1+\delta\pi r_+^2\right)^{1-R/\delta}.
	\label{eq:gibbs_explicit}
\end{align}
This expression may be analyzed parametrically by using the horizon radius $r_+$ as the running variable.

The thermodynamically preferred phase corresponds to the branch with the lowest Gibbs free energy. For pressures below the critical pressure $P_c$, the $G-T_{\rm SM}$ diagram develops a characteristic swallow-tail structure, indicating the coexistence of small and large black-hole phases and signalling a first-order phase transition. As the pressure approaches $P_c$, the swallow-tail shrinks and eventually terminates at the critical point. For $P>P_c$, the swallow-tail disappears and the transition becomes continuous.

\begin{figure}[htbp]
	\centering
	\includegraphics[width=0.45\linewidth]{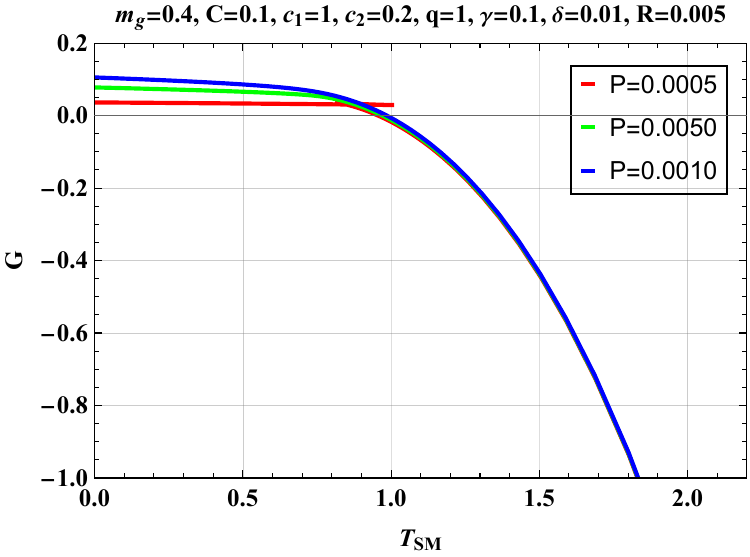}
	\caption{Representative Gibbs free energy $G$ as a function of the Sharma--Mittal corrected temperature $T_{\rm SM}$ for different pressures. The deformation of the Gibbs branches reflects the pressure dependence of the global thermodynamic stability.}
	\label{fig:gibbs_pressures}
\end{figure}

\begin{figure}[htbp]
	\centering
	\includegraphics[width=0.45\linewidth]{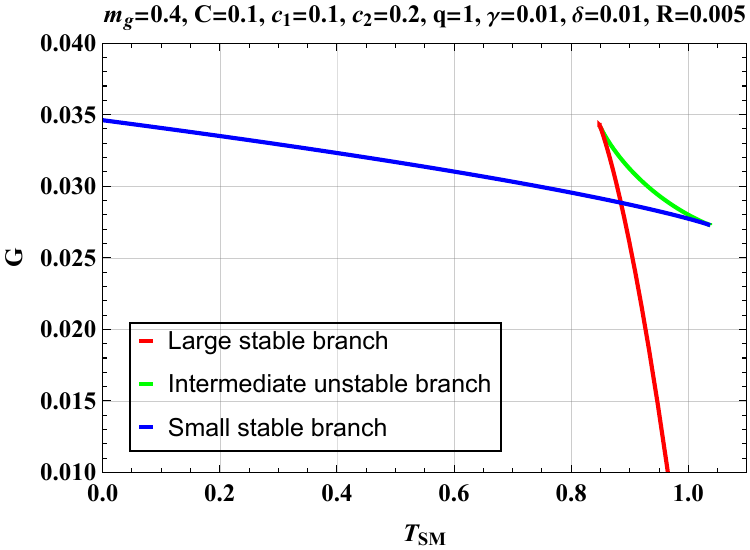}
	\caption{Subcritical Gibbs free energy showing the swallow-tail structure. The small and large black-hole branches correspond to locally stable phases, while the intermediate branch is thermodynamically unstable. The crossing of the stable branches determines the first-order transition temperature.}
	\label{fig:gibbs_swallow_tail}
\end{figure}

The ModMax parameter $\gamma$ modifies the Gibbs structure through the effective charge contribution $q^2e^{-\gamma}$. Increasing $\gamma$ weakens the electromagnetic sector and shifts the transition temperature and pressure. The Sharma--Mittal parameters $(R,\delta)$ alter the thermal contribution $T_{\rm SM}S_{\rm SM}$, thereby affecting the relative stability of competing phases. Meanwhile, the massive-gravity couplings determine the overall shape of the Gibbs landscape and play a decisive role in the emergence of the swallow-tail behaviour.

\subsection{Physical Implications}
\label{subsec:physical_implications_criticality}

The criticality analysis and Gibbs free-energy investigation reveal that the ModMax-dRGT black hole exhibits a rich thermodynamic phase structure analogous to that of a van der Waals fluid. The existence of an inflection point in the equation of state, together with the swallow-tail structure of the Gibbs free energy, confirms the occurrence of a first-order small/large black-hole phase transition terminating at a second-order critical point.

A notable feature of the present model is the distinct role played by the three sectors of the theory. The ModMax nonlinear electrodynamics controls the effective electromagnetic interaction through the exponentially suppressed charge contribution $q^2e^{-\gamma}$, thereby shifting the location of the critical point and modifying the phase boundary. The Sharma--Mittal entropy affects the thermal sector by introducing non-extensive statistical correlations, which influence the stability of competing phases without significantly altering the geometric structure of the critical point. In contrast, the massive-gravity parameters directly enter the equation of state and provide the dominant contribution to the global phase structure.

Taken together, these results demonstrate that the combined effects of nonlinear electrodynamics, massive gravity, and generalized entropy corrections can be efficiently probed through black-hole critical phenomena. Furthermore, the consistency between the criticality analysis and the Gibbs free-energy behaviour establishes a coherent thermodynamic picture in which both local stability and global phase transitions are governed by the interplay between the ModMax field, massive graviton interactions, and non-extensive horizon thermodynamics.
\section{Concluding Remarks}
\label{sec5}

In this paper, we have systematically investigated the Joule-Thomson (JT) expansion of a four-dimensional charged anti-de Sitter (AdS) black hole within the combined framework of ModMax nonlinear electrodynamics and dRGT-like massive gravity. To account for potential non-local statistical correlations and non-extensive effects at the event horizon, our thermodynamic analysis was formulated using the generalized Sharma-Mittal entropy. Operating in the extended phase space, where the cosmological constant is identified with thermodynamic pressure and the black hole mass with enthalpy, we derived exact analytical expressions for the modified Hawking temperature, specific heat, JT coefficient, and the inversion phase boundaries.

Our analysis of the specific heat ($C_P$) revealed that the local thermodynamic stability of the black hole is highly sensitive to both the nonlinear electrodynamic field and the non-extensive entropy parameters. The Sharma-Mittal parameters exhibit competing effects: while an increase in $\delta$ suppresses large-scale thermal fluctuations, increasing $R$ amplifies them, eventually driving the system toward thermodynamic divergences in the strong massive gravity regime. Notably, the ModMax parameter ($\gamma$) acts as a stabilising agent. In regimes of strong massive gravity, where classical Maxwell electrodynamics would yield highly volatile specific heat profiles with multiple phase transitions, the conformal nonlinearities of the ModMax field smooth out these instabilities, establishing a continuous and stable thermodynamic evolution.

The investigation of the isenthalpic throttling process further highlighted the dominant macroscopic role of the ModMax field. We found that the conformal nonlinearities actively shift the JT divergence point to smaller horizon radii, implying that the transition between heating and cooling occurs earlier in the black hole's geometric evolution. When projecting the global inversion phase boundary onto the $T-P$ plane, increasing $\gamma$ significantly shifts the inversion curve downward. Because the cooling region lies above this boundary in standard AdS thermodynamics, this downward shift demonstrates that ModMax nonlinearities effectively expand the physically accessible cooling domain of the black hole during expansion. 

A particularly striking result of our study is the decoupling of microscopic statistical effects from the global macroscopic phase boundary. While the Sharma-Mittal parameters ($\delta$ and $R$) critically dictate the local specific heat and the exact geometric location of the inversion radius, the global inversion curve ($T_i$ vs $P_i$) is virtually immune to their variation. The macroscopic boundary separating the heating and cooling phases is fundamentally governed by the spacetime geometry—specifically the graviton mass parameters—and the ModMax electrodynamic field, rather than the statistical distribution of the horizon microstates. Furthermore, consistent with other AdS models, the inversion curve retains a single, monotonically increasing branch, distinguishing black hole thermodynamics from the dual-branch nature of classical van der Waals fluids.

Furthermore, our investigation into the $P-V$ criticality and global thermodynamic stability confirms that the ModMax-dRGT black hole exhibits a rich phase structure strictly analogous to a van der Waals fluid. By analyzing the inflection points of the geometric equation of state alongside the Gibbs free energy, we established the existence of a first-order small/large black hole phase transition that terminates at a second-order critical point. The physical manifestation of this phase coexistence is explicitly captured by the characteristic swallow-tail structure in the $G-T_{\rm SM}$ diagram for subcritical pressures ($P < P_c$), which smoothly transitions into a continuous phase for $P > P_c$.

Crucially, this critical phase behavior distinctly isolates the respective contributions of the three underlying theoretical frameworks. The ModMax nonlinear electrodynamics controls the effective electromagnetic interaction via an exponentially suppressed charge contribution, shifting the critical point toward smaller horizon radii and actively modifying the phase boundary. The Sharma-Mittal parameters ($\delta$, $R$) introduce non-extensive statistical correlations that influence the thermal sector and the relative stability of competing phases, though they do not strictly alter the geometric structure of the critical point. Ultimately, the massive gravity parameters provide the dominant contribution to the global phase landscape, dictating the overall thermodynamic geometry and the emergence of the swallow-tail behavior. Collectively, these results reinforce the utility of AdS black holes as robust theoretical laboratories, demonstrating how the complex interplay of nonlinear electrodynamics, massive graviton interactions, and non-extensive thermodynamics can be elegantly probed through critical phase phenomena.

\section*{Acknowledgments}
DJG acknowledges the contribution of the COST Action CA21136  -- ``Addressing observational tensions in cosmology with systematics and fundamental physics (CosmoVerse)". Also, HH is grateful to Excellence project FoS UHK 2203/2025-2026 for the financial support.

\section*{Declaration of competing interest}
The authors declare that they have no known competing financial interests or personal relationships that could have appeared to influence the work reported in this manuscript.

\section*{Data Availability Statement}
There are no new data associated with this article.

\bibliography{refs}
\end{document}